%% file: main.tex
\documentclass{article}%
\usepackage{amsmath,amsfonts}
\usepackage{algorithmic}
\usepackage{algorithm}
\usepackage{array}
\usepackage[letterpaper,top=2cm,bottom=2cm,left=2cm,right=2cm,marginparwidth=1.75cm]{geometry}
\usepackage[caption=false,font=normalsize,labelfont=sf,textfont=sf]{subfig}
\usepackage{textcomp}
\usepackage{stfloats}
\usepackage{url}
\usepackage{verbatim}
\usepackage{graphicx}
\usepackage[default,scale=0.9]{opensans}
\usepackage[T1]{fontenc}
\usepackage[utf8]{inputenc}
\usepackage{caption}
\usepackage{float}
\usepackage[
backend=biber,
style=numeric,
sorting=none,
]{biblatex}
\usepackage{fancyhdr}
\date{}

\title{Automated Consolidation Of Dispersed DSO Data Resources\\Into A Consistent Digital Twin Of The Distribution Grid}
\author{Fabian B\"ohm$^{1,2}$$^*$, Benjamin Sawicki$^2$,Claudio Stucki$^1$,Pulkit Nahata$^2$,Gabriela Hug$^2$,Lukas Ortmann$^1$ } 
\begin{document}
\maketitle
 \begin{center}

    \textit{$^1$Eastern Switzerland University of Applied Sciences (OST), Rapperswil-Jona, Switzerland \\
    $^2$ETH Z\"urich, Z\"urich, Switzerland  \\
    $^*$fboehm@ethz.ch}

 \end{center}
 \begin{center}
  \end{center}
\pagestyle{fancy}
\fancypagestyle{plain}{
\fancyhf{} 
\fancyhead[C]{\small This paper is a preprint of a 
paper accepted by IET conference proceedings and is subject to Institution of Engineering and Technology Copyright. When the final version is published, the copy of record will be available at IET Digital Library}%
\fancyfoot[C]{\small\thepage}
}
\fancyhf{}
\fancyhead[C]{\small Published in IET Conference Proceedings, Volume 2026 Issue 18, cite as https://doi.org/10.1049/icp.2026.3744}
\fancyfoot[C]{\small\thepage}

\section*{Abstract}
The energy transition causes a growing number of distributed flexible energy resources in the low voltage grid. To utilize that flexibility, automation and control algorithms are needed, which typically are developed in theoretical research. We build a cyberphysical algorithm test bed to bridge the gap between theoretical research and practical application in cooperation with a local DSO in Walenstadt, Switzerland. To obtain suitable monitoring and data analysis capabilities from their existing data landscape, we propose tool-independent data harmonization measures in preparation for a digital twin of the grid. For data extraction from high-security network zones we developed a secure pipeline that implements a logical data diode ensuring one-way traffic only. Furthermore, we established a test community of voluntary prosumer households, whose flexible assets were made accessible through a common backend, giving us actuation capabilities. The presented publish-subscribe architecture is not only suitable to integrate various data sources into a central data collection, but also to manage access privileges of different stakeholders. This could support unbundling of an organization’s data flows and the implementation of data space concepts.

\section{Introduction}
The increasing number of flexible asset in the low voltage (LV) distribution grid on end customer level such as heat pumps (HP), electric vehicles (EV), stationary batteries and photovoltaic (PV) systems imposes new challenges and opportunities for grid operation, energy markets and business models \cite{2023_iea_tracking}. Advanced control algorithms enable automated operation of these distributed resources and make their flexibility available for different applications \cite{2022_rser_dettore}. To bridge the gap between theoretical research and practical application in today's distribution grid we build a cyberphysical algorithm and controls test bed.

Other initiatives strive to achieve this goal by refining an initially formulated abstract concept (for example European energy data spaces proposed by OMEGA-X, EDDIE and others \cite{2026_dgconnect_eudataspaces}, or conceptual Smart Grid platforms \cite{2024_cired_karmann}, \cite{2024_cired_fiedler}) into specific requirements that can be implemented in practice. Complementary to this top-down approach, we identify the steps and technologies required to achieve the same goal starting from a DSO's day-to-day operation and legacy software systems in use in a bottom-up manner.
Therefore existing assets and systems are integrated into the proposed framework with minimal overhead while laying the foundation for a wide range of research opportunities.
Like many electricity utilities, our project partner manages their grid related data in use-case-specific software tools and documents, that are optimized for day-to-day operation rather than research and analysis. Data from different sources must be linked, transformed into machine-readable and preferably open formats, and kept up to date automatically. Besides data harmonization, real-time monitoring and command transfer are essential to utilize flexible assets in closed-loop control schemes. 

The remainder of this paper is structured as follows: In section \ref{sec_ex_data_landscape} the available data sources at the DSO are analysed and key steps for data harmonization are identified. In section \ref{sec_test_comm} we describe how end-customer flexibility was integrated. After presenting our solution for automated data extraction from secure network environments in section \ref{sec_data_extr}, the paper is concluded in section \ref{sec_conc}.

\input{01_existingdatalandscape}
\input{02_testcommunity}
\input{03_dataextraction}
\input{04_conclusion}
\section{Acknowledgements}
This work was supported by the NCCR Automation, a National Centre of Competence in Research, funded by the Swiss National Science Foundation (grant number 51NF40\_225155).

\printbibliography
\end{document}

%% file: 01_existingdatalandscape.tex
\section{Existing data landscape} \label{sec_ex_data_landscape}
The project's main data sources are
\begin{enumerate}
    \item \textbf{Substation measurements}: 46 power quality sensors measure electrical quantities at almost all MV/LV transformers of the grid. The results are collected by a server within a protected OT (Operational Technology) network section.
    \item \textbf{Test community}: 18 prosumer households deliver behind-the-meter measurements of various flexible assets and inflexible consumption.
    \item \textbf{Grid related data}: Additional documents, models and reports from the grid operator's daily operation.
\end{enumerate}
In the remainder of this section the identified types of data sources and their respective properties are presented, and recommendations for data harmonisation across data silos are made.
\subsection{Stream-based and artifact-based data sources}
The available data sources can be categorized as either artifact-based or stream-based sources. 
Artifacts are snapshot-like files representing a fixed system state, either in human readable Excel and pdf documents or in specialized formats such as GIS shape files or a proprietary grid simulation model. 

In some cases, the information contained in an artifact is structured according to a specific data model (a representation of the things of significance to an enterprise, the relations among those things and the underlying structure of the enterprise’s data \cite{1996_dorset_hay}) that is not provided explicitly in form of an XML schema or similar. Other artifacts like reports and manually curated documents have no set structure at all. 
Only when significant changes occur, they are updated with a lot of manual effort and often with the involvement of third-party contractors. We retrieve artifact-based data manually via secure online file exchange channels.

Stream-based data sources on the other hand continuously collect timestamped data in tabular form, e.g. evolving measurement results and status values from assets in the field. New entries are appended to database tables, either periodically at the sampling interval or in a consolidated manner, and conform to a manufacturer specific data model. We retrieve the stream-based test community data and substation measurement data automatically via the respective REST APIs: The test community households are equipped with home energy management systems (HEMS) that transfer data to the HEMS supplier's cloud, from where an authenticated user can retrieve the data over the public internet. The API of the substation measurements server on the other hand is located within a closed OT network. Both systems provide data in a machine-readable format. To build a comprehensive data collection incorporating data from all available sources at the research server, a logical data diode pipeline for secure substation measurement extraction from the utility's private network was developed. 

\subsection{Ambiguity in asset naming}
\begin{figure}[H]
\centering
\includegraphics[width=0.9\linewidth]{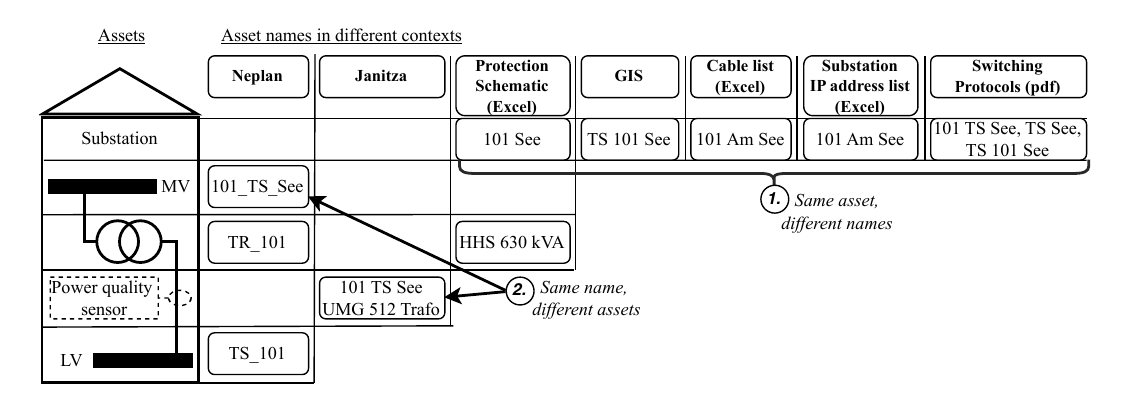}
\caption{Ambiguity between assets and their names}
\label{fig_entities_names}
\end{figure}

Because the existing data is primarily used by the utility's technicians, human-readable identifiers are used to refer to physical assets (e.g. a substation) in all data sources. Across data sources an identifier contains identical information, but comes in different formats (shown as \textit{same asset, different names} in \ref{fig_entities_names}), depending on creator, data source and context. 
Human users can match the identifiers to a unique entity with ease, but for automated data processing unstructured ID strings pose a significant challenge: Extraction of the contained information from arbitrarily formatted identifiers that is robust towards all thinkable exceptions would require a massive implementation effort, with questionable reliability. A global naming convention for entity identifiers across all data sources is independent of the software tools in use and overcomes the presented challenge.
After identifier harmonization new challenges arise: The same identifier across different data sources can refer to different things, e.g. a substation (building), a bus bar (node in the electrical network), a transformer, or a power quality sensor. This occurs because the utility's technicians often refer to components within a substation using the substation's name, but no explanation of the relations among the substation's components is provided. One can only conclude the intended component from a document's context. The second ambiguity \textit{same name, different asset} in figure \ref{fig_entities_names} illustrates this situation, where measurements from the low voltage busbar of a transformer are incorrectly assigned to the transformer's high voltage busbar in the simulation model. 

\begin{figure}[H]
\centering
\includegraphics[width=0.5\linewidth]{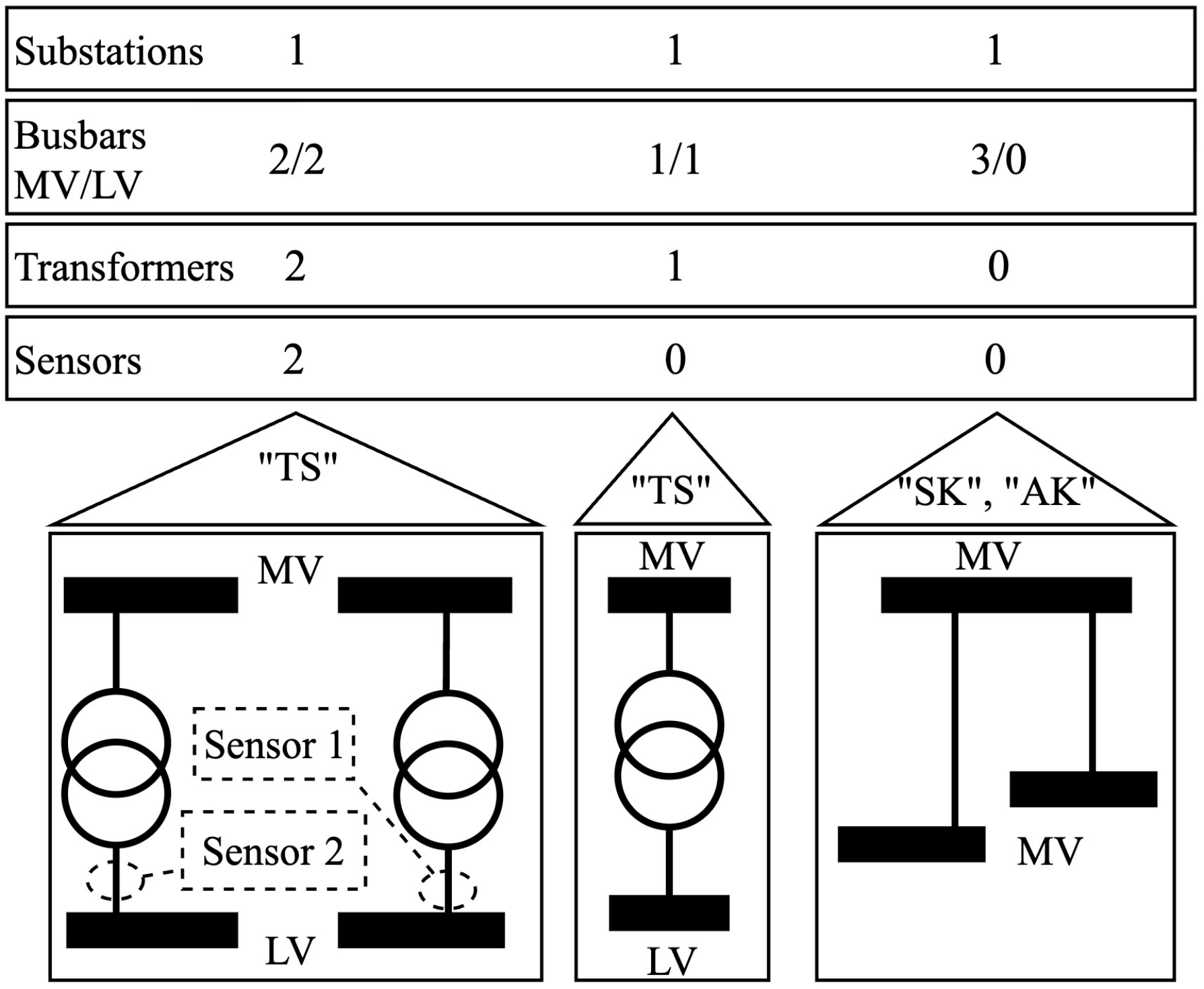}
\caption{Different configurations of substation components}
\label{fig_substation_types}
\end{figure}

The different substation configurations shown in figure \ref{fig_substation_types} can be found in the distribution grid, therefore naming conventions should take the individual components' naming into consideration.
Manual tasks in daily operation are usually conducted within a single data source, therefore this ambiguity is not noticeable. The global naming scheme must differentiate between assets located at the same substation explicitly. Time information in file names, reports and databases should comply to the ISO 8601 standard to avoid ambiguity. 

%% file: 02_testcommunity.tex
\section{Test community} \label{sec_test_comm}
To build a smart grid test bed at a high technology readiness level (TRL 6-7), the existing sensing and actuation assets must become remotely accessible for monitoring and control. The established flexible asset pool comprises different device types from different generations of various manufacturers and lacks standardized communication interfaces and data models. Home energy management systems, an emerging family of devices, hide that complexity behind a uniform communication gateway for all flexible assets in a prosumer household.

\subsection{Test community selection process} \label{sec_tcom_sel}
We assessed the suitability of existing assets in our community based on their communication capabilities and maximum power ratings. 
Based on knowledge of the available devices and their compatibility, we conducted a cost-benefit analysis, considering the criteria in \ref{tab_benefit_analysis}. Note that \textit{full} measurement set or controllability indicates that continuous quantities are available for monitoring and control, whereas \textit{restricted} indicates that less values (e.g. only discrete status values) are available.
\clearpage
\begin{center}
\captionof{table}{Decision criteria for flexible assets}
\begin{tabular*}{0.45\textwidth}{|p{0.10\textwidth}|p{0.3\textwidth}|}
 \hline
 \textbf{Device} & \textbf{Criteria / Compatibility} \\  \hline
 PV-System & Installed peak power \\  \hline
 Inverter & measurement-only or controllability of active and reactive power \\  \hline
 Smart Meter & full or restricted measurement set, \newline additional hardware required\\  \hline
 Battery & full or restricted controllability, \newline  additional hardware required  \\ \hline
 EV-Charger & full or restricted controllability, \newline  additional hardware required \\  \hline
 EV &  full or restricted meaurements \\  \hline
 Heat pump & full or SG-ready, \newline   additional hardware required \\  \hline
 Boiler & full or restricted \\  \hline

\end{tabular*}
\label{tab_benefit_analysis}

\end{center}

The maximum of points are given to devices that are compatible to the chosen HEMS and offer a digital communication interface (e.g. ModbusTCP), allowing two operations essential for a smart grid test bed: First, to read device information (e.g. measurements, status values), and second, to continuously alter the devices' operation (e.g. target power injection or temperature). %
We observe that a single device can offer multiple communication interfaces with different functionalities, e.g. an inverter's web interface offers read-only access to measurements, while its ModbusTCP interface allows not only to read measurements but also to set new target values for controllable quantities.
Lower scores are awarded to devices if they offer restricted communication interfaces (such as proprietary interfaces), discrete actuation capabilities only (e.g. relay contacts instead of set-points) or if additional hardware (e.g. a network card) is required for remote accessibility.
Zero points are awarded for devices that are incompatible to the chosen HEMS and devices that operate completely autonomously with no communication interface at all, concerning mainly older devices owned by early adopters.

To account for varying importance of the presented criteria, the overall rating was calculated as a weighted sum of these scores. Afterwards the identified benefits were set into relation to the expected installation costs.
Based on this evaluation, we ordered the participating houses with respect to their suitability for the project and installed HEMS in 18 houses.

\subsection{HEMS: Make or Buy} \label{makeorbuyhems}
We conducted a market research to identify a suitable HEMS considering compatible communication interfaces, total lifetime cost, automated energy optimization and willingness of the HEMS manufacturer to cooperate in our project.
Apart from the unified communication interface for all devices, we identified further reasons to use an off-the-shelf HEMS: The HEMS installation guide provided by the manufacturer simplifies the rollout process for the electricians. After installation, the HEMS performs continuous self-consumption optimization of the household utilizing the available flexible assets. At the same time, customers can use the associated mobile and web app to monitor their energy production and consumption. We learned that some test community members became curious about their energy consumption and started to optimize their behaviour accordingly, while others rely on the optimization carried out by the HEMS. A raised awareness for energy consumption can positively influence a customer's energy consumption, but \cite{2024_scas_cellina} states that the positive effect observed declines after a year. 

\subsection{Test community data retrieval}
Complementary to end customer apps the HEMS manufacturer cloud provides a REST API which allows to poll data automatically, namely continuously collected measurement results and state variables. This data conforms to a vendor-specific schema and is stored in relational database tables. On the contrary, static data describing the physical properties and the operating range of assets changes rarely and is not retrievable from the REST API. Therefore we collected the parameters in a manual survey among potential test community members prior to the selection in \ref{sec_tcom_sel}.

\subsection{Expanding the available flexibility}
As a first step, we use the collected data to assess and improve the performance of the flexible assets. 
There are many flexible assets in Walenstadt, and our remote access is limited to the small group in the test community. Even within this remotely accessible group of devices, different aspects lowered the actually available flexibility: Evaluating the collected data, we identified broken components (e.g. a heatpump valve), incompatible configurations (e.g. external commands were ignored), non-operative devices (e.g. a battery that was offline over months due to a fault) and communication interface bugs. These faults do not impose security risks but result in less efficient operation. Continuous operation monitoring and evaluation are essential to maximize the available flexibility, even if the components are designed to operate autonomously.

Currently our platform supports only one HEMS manufacturer, but with a standardized HEMS API devices from other manufacturers could be integrated easily. The functional profiles developed by the initiative SmartGridReady CH \cite{2026_sgr_ch} aim to unlock such vendor-independent interoperability.

%% file: 03_dataextraction.tex
\section{Data Extraction Pipeline}\label{sec_data_extr}
The heterogeneity of data sources, combined with strict cybersecurity requirements in OT environments, necessitated the design of a secure-by-design data extraction architecture. A central constraint was the preservation of network segregation and segmentation principles commonly applied in critical infrastructure: Data originating from high-security OT zones must not introduce new attack surfaces or bidirectional connectivity that could compromise system integrity. At the same time, the architecture had to support scalable, reliable, and extensible data flows across institutional and organizational boundaries.

Classical site-to-site VPN solutions were deliberately avoided. While VPNs provide encrypted network tunnels, they effectively extend trust boundaries and expose internal services to external networks. Even modern zero-trust VPN approaches primarily secure transport channels but still rely on persistent network connectivity and endpoint exposure, without offering inherent messaging, buffering, or decoupling mechanisms \cite{2025_ibmxforce_mcmillen}. For a research-driven multi-stakeholder ecosystem with evolving participants, such tightly coupled network integration was considered both operationally complex and misaligned with security-by-design principles.

Instead, we adopted a publish-subscribe (pub-sub) messaging architecture as the backbone of the data extraction pipeline. Pub-sub systems enable producers within secure network zones to initiate outbound connections over Transport Layer Security (TLS) sessions, thereby maintaining the unidirectional communication principle often referred to as a “logical data diode” \cite{2010_csiir_okhravi}. Data sources publish messages to well-defined topics or queues hosted on a broker infrastructure located at our central research data server, while consumers subscribe independently to that broker according to their access privileges and use cases.

Early architectural evaluation focused on mature, high-throughput messaging platforms, particularly Apache Kafka and RabbitMQ. Although the functional gap between modern versions of both systems has narrowed considerably, RabbitMQ was selected. The pub-sub architecture provides several key properties essential for the Grid2050 data ecosystem: %
\begin{enumerate}
    \item \textbf{Spatial decoupling} is achieved by independently operating publishers and subscribers with the message broker acting as a buffer in between. The presented pub-sub architecture furthermore includes modern messaging guarantees such as at-least-once delivery, ensuring complete data transmission despite transient network outages. %
    \item \textbf{Temporal decoupling} allows non-real-time processing without data loss. While many data sources provide near real-time measurements, downstream systems such as databases, analytics platforms, or control experiment environments may be temporarily unavailable due to maintenance. During such periods, messages accumulate in the queues and are processed once consumers resume operation, eliminating tight synchronization requirements. The queues are near empty and extremely fast under normal operating conditions.
    \item \textbf{Security} is enforced at multiple layers. Communication channels over the public internet are encrypted using TLS 1.3, and strong mutual authentication is implemented via X.509 certificates. Authorization is applied per queue and exchange, enabling strict segregation of data flows according to organizational roles and research use cases. %
    \item \textbf{Availability and fault tolerance} can be incrementally enhanced. While initial deployment relied on classic queues, RabbitMQ supports quorum queues that replicate message streams across multiple broker nodes using consensus protocols. We also intend to add redundancy at different stages of the pipeline in the future.%
\end{enumerate}

The command-constrained interaction model enabled by pub-sub communication introduces an additional security mechanism. Nodes in high-security zones subscribe to specific command topics, allowing external systems to request certain actions from a predefined set of actions. Unlike VPN-based remote access, only structured messages are delivered and processed by explicitly implemented handlers, substantially reducing attack surface.

Finally, redundancy and scalability are inherent to the pub-sub messaging architecture. Multiple producers can publish equivalent data streams, and multiple consumers can independently process the same information without introducing coordination complexity. This enables parallel research data streams, replication of critical data extraction paths, parallel tests of different database systems and future expansion of the integrated data sources without architectural redesign.

With this messaging infrastructure in place, the pipeline can integrate a wide range of heterogeneous data sources, including SCADA and OT network measurements, residential prosumer systems, weather sensors and cameras, and external online services. All streams converge towards a central data platform, where a primary subscriber saves the content of the incoming streams to a unified digital twin database, serving as the system’s single source of truth. Downstream subscribers — distributed across research groups and partner institutions — can consume live updates for visualization dashboards, control algorithm testing, forecasting model training, and simulation experiments. Each platform evolves independently while remaining synchronized through the shared event stream.

This event-driven data extraction pipeline thus forms the foundational layer of the Grid2050 digital twin, enabling secure cross-domain data integration, operational resilience, and scalable research collaboration without compromising critical infrastructure security principles.

%% file: 04_conclusion.tex
\section{Conclusion} \label{sec_conc}
While building a cyberphysical algorithm test bed in the distribution grid, necessary steps for modern grid operation were identified. As a first software-independent step we propose global naming conventions for relevant assets, taking into consideration relations between entities and components. 
The existing grid related data sources are classified as either stream-based or artifact-based. Artifacts come in heterogeneous, unstructured formats and are updated irregularly resulting from manual business processes. Adjusting the business processes to generate artifacts in structured, machine-readable open-source formats simplifies the integration of those data sources into a digital twin significantly.
In addition to the unified data collection built on DSO resources, remote monitoring and actuation of end customer flexibility is realised by using HEMS to make flexible assets remotely accessible. The implemented common cloud backend for the test community is a temporary solution which can be replaced by a future vendor-independent HEMS API. Using the collected test community data we identified and fixed various problems that reduced the available flexibility.
The publish-subscribe infrastructure decouples the digital twin of the grid from distributed data sources, distributed research institutions, private households, and algorithmic controllers operating in non-monopolistic domains. The event-driven model further provides a practical mechanism to bridge traditionally isolated OT environments with modern IT analytics platforms without extending trust boundaries or compromising cybersecurity principles. Advanced algorithms can thus be deployed across organizational domains while preserving strict network segregation. Consequently, Grid2050 should not be understood as a smart metering or smart home demonstrator, but rather as an interconnected cyber-physical experimentation environment for an automated, distributed energy system. 

This modular approach positions the platform as an integration layer toward emerging Common European Data Spaces, where heterogeneous standards currently prevail. Future work could also focus on the integration of data consumers with different access privileges into the publish-subscribe architecture to selectively exchange required datasets from a central digital twin database, which could support the organizational unbundling of European utility companies.